\documentclass[aps,prl, a4paper,twocolumn,10pt,superscriptaddress]{revtex4}

\usepackage{amsmath}
\usepackage{makeidx}
\usepackage{amsfonts}
\usepackage[utf8]{inputenc}
\usepackage[usenames,dvipsnames]{pstricks}
\usepackage{epsfig}
\usepackage{subfigure}
\usepackage{pst-grad} 
\usepackage{pst-plot} 
\usepackage[colorlinks,hyperindex]{hyperref}
\usepackage{epstopdf}

\hypersetup
	{
		colorlinks,%
		citecolor=black,%
		linkcolor=black,%
		urlcolor=black,%
	}

\makeindex    

\begin{document}

\title{Interferences in quantum eraser reveal geometric phases in modular and weak values}

\author{Mirko Cormann}
\email{mirko.cormann@unamur.be}
\affiliation{Research Centre in Physics of Matter and Radiation (PMR), Department of Physics, University of Namur, 61 Rue de Bruxelles, B-5000 Namur, Belgium}
\affiliation{Namur Center for Complex Systems (naXys)}
\author{Mathilde Remy}
\affiliation{Research Centre in Physics of Matter and Radiation (PMR), Department of Physics, University of Namur, 61 Rue de Bruxelles, B-5000 Namur, Belgium}
\author{Branko Kolaric}
\affiliation{Research Centre in Physics of Matter and Radiation (PMR), Department of Physics, University of Namur, 61 Rue de Bruxelles, B-5000 Namur, Belgium}
\author{Yves Caudano}
\email{yves.caudano@unamur.be}
\affiliation{Research Centre in Physics of Matter and Radiation (PMR), Department of Physics, University of Namur, 61 Rue de Bruxelles, B-5000 Namur, Belgium}
\affiliation{Namur Center for Complex Systems (naXys)}

\date{\today}

\begin{abstract}
In this letter, we present a new procedure to determine completely the complex modular values of arbitrary observables of pre- and post-selected ensembles, which works experimentally for all measurement strengths and all post-selected states. This procedure allows us to discuss the physics of modular and weak values in interferometric experiments involving a qubit meter. We determine both the modulus and the argument of the modular value for any measurement strength in a single step, by controlling simultaneously the visibility and the phase in a quantum eraser interference experiment. Modular and weak values are closely related. Using entangled qubits for the probed and meter systems, we show that the phase of the modular and weak values has a topological origin. This phase is completely defined by the intrinsic physical properties of the probed system and its time evolution. The physical significance of this phase can thus be used to evaluate the quantumness of weak values.
\end{abstract}




\maketitle

In 1988, Aharonov, Albert, and Vaidman (AAV) introduced the weak value of a quantum observable $\hat{A}$ from an extension of the von Neumann measurement scheme \cite{Aharonov (1988)}. They pointed out that the result of a measurement involving a weak coupling between a meter and the observable $\hat{A}$ of a system with a pre-selected initial state $|\psi_{i}\rangle$, and a post-selected final state $|\psi_{f}\rangle$ depends directly on the weak value:
\begin{equation} \label{eq:WeakValueDefinition}
A_{w}=\frac{\langle\psi_{f}|\hat{A}|\psi_{i}\rangle}{\langle\psi_{f}|\psi_{i}\rangle}\:,
\end{equation}
an unbounded complex number. In particular, they showed that the shift of the average detected position due to post-selection is proportional to the real part of the weak value. Since for weak measurements in the absence of post-selection, this shift is proportional to the average of the observable $\langle\psi_{i}|\hat{A}|\psi_{i}\rangle / \langle\psi_{i}|\psi_{i}\rangle$, a direct but bold physical interpretation of the weak value assumes it represents somehow the average of $\hat{A}$ in the pre- and post-selected ensemble. They also related the imaginary part of the weak value to the shift of the average impulsion. Beside the AAV approach, weak values may also appear using a meter strongly coupled to the observable $\hat{A}$ \cite{Kofman (2012),Brun (2008),Wu (2009),Shikano (2011),Hofmann (2013)-1,Iinuma (2011)}. In these instances, the effective weak interaction is achieved by selecting particular initial states of the meter system, so that the probability of actually measuring $\hat{A}$ is low and the probed system is left unperturbed most of the time. Therefore, both methods transform the standard von Neumann procedure to a weak measurement with a high incertitude.

Weak values and weak measurements proved useful in many fields of physics and chemistry \cite{Hosten (2008),Dixon (2009),Starling (2009),Tang (2010),Rhee (2009), Solli (2004),Brunner (2004),Resch (2004),Lundeen (2009), Kocsis (2011),Lundeen (2011),Salvail (2013),Lundeen (2012)}. Nevertheless, the proper physical interpretation of weak values remains highly debated. For example, weak values were used to develop a time-symmetrized approach to standard quantum theory, the two-state vector formalism \cite{Aharonov (2008)}, where they appear as purely quantum objects. Oppositely, a purely classical view of the occurrence of unbounded, real weak values -- and possibly of complex ones -- was proposed recently \cite{Ferrie (2014)} (which is criticizable though \cite{Brodutch (2015),Vaidman (2014),Dressel (2015)}).

In this letter, we uncover a physical interpretation of complex weak values in terms of their polar representation (modulus and argument), which provides evidence for their quantumness. We devised an interferometric procedure to measure and discuss complex weak values in their polar representation instead of the usually determined real or imaginary part. Our procedure relies essentially on a joint phase and visibility measurement in a quantum interferometer where the meter system acts as a quantum eraser. Using simple cases exploiting entangled qubits, we relate the argument of the weak value to topological phases defined completely by the probed system states involved in the weak measurement. Additionally, our procedure works in conditions where the usual weak measurement procedure fails completely: (\mbox{I}) for arbitrary measurement strengths (i. e. including strong measurements) and (\mbox{II}) for orthogonal and nearly orthogonal initial and final probe states. It proceeds by optimizing the interference phase to measure simultaneously the modulus and the argument of the weak value in a single step.

Formally, our procedure implements a quantum controlled evolution, in which an arbitrary quantum system $|\psi_{i}\rangle$, the probe,  interacts with a qubit meter via the  quantum gate (Fig. \ref{fig:bloch representation}.a): 
\begin{equation}
\hat{U}_{GATE}=\hat{\Pi}_{r}\otimes\hat{I}+e^{i \delta}\,\hat{\Pi}_{-r}\otimes\hat{U}_{A}\;,\label{eq:controlled measurement interaction}
\end{equation}where $\hat{\Pi}_{\pm r}$ are projectors acting on the meter and $\delta$ is a phase factor first supposed to be null. The transformation $\hat{U}_{A}=e^{-ig\,\hat{A}}$ is expressed in terms of a time independent Hermitian operator $\hat{A}$ and an arbitrary coupling strength $g$, defined by the integral $g=\int g\!\left(t\right)\, dt$ \cite{Comment Hint}. After the gate interaction, the spin observable $\hat{\sigma}_{q}$ of the meter is measured. According to the final meter state, the information about whether the transformation $\hat{U}_{A}$ was applied on the probe can be preserved or erased, completely or partially. Finally, a projective measurement of the probe system post-selects the vector state $|\psi_{f}\rangle$.

The average $\overline{\sigma}_{q}^{m}$ of the meter observable for a given pre- and post-selected sub-ensemble of the probe system is then:
\begin{equation}
\overline{\sigma}_{q}^{m}=2 P_{m}\frac{\left(\overrightarrow{m}.\overrightarrow{q}\right)\: \operatorname{\Re \mathfrak{e}} A_{m} +\left[ \left(\overrightarrow{r}\times\overrightarrow{m}\right).\overrightarrow{q}\right] \:\operatorname{\Im \mathfrak{m}} A_{m} }{\left(1+P_{m}\,\overrightarrow{r}.\overrightarrow{m}\right)\:+\left(1-P_{m}\, \overrightarrow{r} . \overrightarrow{m}\right)\: \left|A_{m}\right|^{2}}\:.\label{eq:QSMeanValue}
\end{equation}
In this expression, the normalized vectors $\overrightarrow{m}$, $\overrightarrow{q}$ and $\overrightarrow{r}$ are the directions on the meter Bloch sphere determining the initial $| m \rangle$ and final $| q \rangle$ meter states as well as the projector state $| r \rangle$ controlling the interaction, respectively. The direction of $\vec{q}$ was chosen orthogonal to $\vec{r}$ to select maximally interfering pathways through the meter measurement: then, the gate action appears as a superposition of having applied both $\hat{U}_{A}$ and $\hat{I}$, and all information about the gate action is lost (quantum eraser condition). The parameter $P_{m}$  characterizes the purity of the initial meter state, ranging from $1$ for pure states to $0$ for a maximally mixed state. $A_{m}$ is defined as the modular value of the probe observable $\hat{A}$ for the pre- and post-selected sub-ensemble: 
\begin{equation}
A_{m}=\frac{\langle\psi_{f}|e^{-ig\hat{A}}|\psi_{i}\rangle}{\langle\psi_{f}|\psi_{i}\rangle}\: .\label{eq:ModularValue}
\end{equation}
It appears from the action of projectors in (\ref{eq:controlled measurement interaction}). Modular values were not often reported as such in the literature because they are directly related to weak values in the usual weak approximation limit  for small coupling strengths, through a first order polynomial development in $g$: $A_{m}=1-ig\ A_{w}+\operatorname{o}\!\left(g^{2}\right)$. Nevertheless, they characterize all projective couplings between the probe and meter systems, where they generalize weak values in a non-perturbative way. They typically describe quantum-gate type interactions  \cite{Kedem (2010)} and quantum interference experiments \cite{Sponar (2015), Tollaksen (2010), Denkmayr (2014)}, but appear also in photon trajectory measurements \cite{Kocsis (2011)} for example. In the following, we relate the physical interpretation of modular values to the visibility and the phase of interferometric experiments.

In our procedure, the interaction strength is not determined by $g$. Instead, it reflects the probability of the application of $\hat{U}_{A}$ by the quantum gate, which is controlled by the measurement strength $\theta=\arccos\left(\overrightarrow{m}.\overrightarrow{r}\right)$, with $\theta\in\left[0,\pi \right]$. By choosing particular final meter states $\vec{q}_{Re}$ and $\vec{q}_{Im}$ (constrained by $\vec{q}.\vec{r}=0$), the real and imaginary parts of the modular value are determined from the average meter observable $\overline{\sigma}_{q}^{m}$, respectively \cite{Supplemental Material Meter Configurations}. For small measurement strengths  $\theta\approx 0$ when the purity  $P_{m}$ is close to one, we obtain the modular value according to the standard approximations of weak measurements:
\begin{equation}
\operatorname{\Re \mathfrak{e}} A_{m} \approx\frac{1}{\theta}\,\overline{\sigma}^{m}_{q_{Re}}\:\:\:\:\: \operatorname{\Im \mathfrak{m}}  A_{m} \approx\frac{1}{\theta}\,\overline{\sigma}^{m}_{q_{Im}}\:,\label{eq:realpartmodular variable}
\end{equation}
where the weak measurement approximation effectively removes the nonlinear dependence of equation (\ref{eq:QSMeanValue}) on the modular value modulus (see denominator).

For an arbitrary measurement strength, we seek instead to measure the modular value in its polar form to assess directly its modulus $\left|A_{m}\right|$ and argument $\varphi=\arg A_{m}$. We introduce an additional unitary transformation $\hat{R}_{\xi}$ in the meter path. It creates a relative phase shift $\xi$ between the orthogonal states $|r\rangle$ and $|-r\rangle$ that is effectively equivalent to a rotation of the modular value in the complex plane. When the phase shift compensates precisely the argument of the modular value (i. e. when $\xi=\varphi$), this rotation aligns the modular value with the real axis. Choosing the meter configuration $\vec{q}_{Re}$ that picks the real part of the modular value provides now its full modulus, while its argument is equal to the introduced phase shift. In practice, our procedure implements a quantum interferometer exploiting entanglement to measure the two quantities concurrently. Indeed, the expression for the joint probability outcome $P{}_{joint}$ of the meter and the probe measurements is proportional to: 
\begin{equation}
P_{joint}\propto1+V\, \cos\left(\varphi-\xi\right)\:,\label{eq:jointprobaQuantumeraser}
\end{equation}
typical of an interference phenomenon, where $V$ represents the visibility and $\varphi-\xi$ the phase.  Experimentally, the visibility is determined by measuring the maximum and the minimum of the joint probability, denoted by $P_{max}$ and $P_{min}$, respectively: 
\begin{equation}
V=\frac{P_{max}-P_{min}}{P_{max}+P_{min}}\:.\label{eq:visibility}
\end{equation}
When the phase introduced by $\hat{R}_{\xi}$ equals the argument of the modular value, the maximum of the joint probability is obtained for the meter vector $\vec{q}_{Re}$, while its minimum is obtained for the orthogonal state $-\vec{q}_{Re}$. The two situations correspond to constructive and destructive interferences in the joint measurement, respectively. The visibility depends on the coupling strength and the modular value modulus:
\begin{equation}
V=\frac{2\,P_{m}\, \tan\left(\frac{\theta}{2}\right)}{C_{\theta+\pi}+C_{\theta}\:\tan^{2}\left(\frac{\theta}{2}\right) \left|A_{m}\right|^{2}}\,\left|A_{m}\right|\:,
\end{equation}
with coefficients $C_\epsilon$ defined by:
\begin{equation}
C_{\epsilon}=\frac{1+P_{m}}{2}+\frac{1-P_{m}}{2}\, \cot^{2}\frac{\epsilon}{2}\,.\label{eq:meter coeff}
\end{equation} 
This quadratic equation provides two solutions for the modular value modulus:
\begin{equation}
\left|A_{m}\right|_{\pm}= \frac{1\pm\sqrt{1-C_{\theta}\, C_{\theta+\pi}\, P_m^{-2} V^{2}}}{C_{\theta}\:\tan\left(\frac{\theta}{2}\right)\, V} P_m \:.\label{eq:modular value relation}
\end{equation}
The solution $\left|A_{m}\right|_{-}$ corresponds to $\left|A_{m}\right|$, if the condition:
\begin{equation}
\tan^{2}\left(\frac{\theta}{2}\right)\frac{C_{\theta}}{C_{\theta+\pi}}\, \left|A_{m}\right|^{2}\leq1\,\label{eq:WeakMeasurCond}
\end{equation}
is verified, and $\left|A_{m}\right|_{+}=\left|A_{m}\right|$ otherwise. Together, they provide the characterization of the modular value modulus for an arbitrary coupling strength. It is directly related to the visibility. In particular, the weak measurement approximation gives $\left|A_{m}\right|\approx V / \theta$, similarly to equation (\ref{eq:realpartmodular variable}). In this expression of the modular value, the visibility plays the same role than the pointer shift in weak values. This shows a strong connection between modular values and weak interferometric experiments.

\begin{figure}[t]
	\begin{center}
			\includegraphics[width=0.48\textwidth]{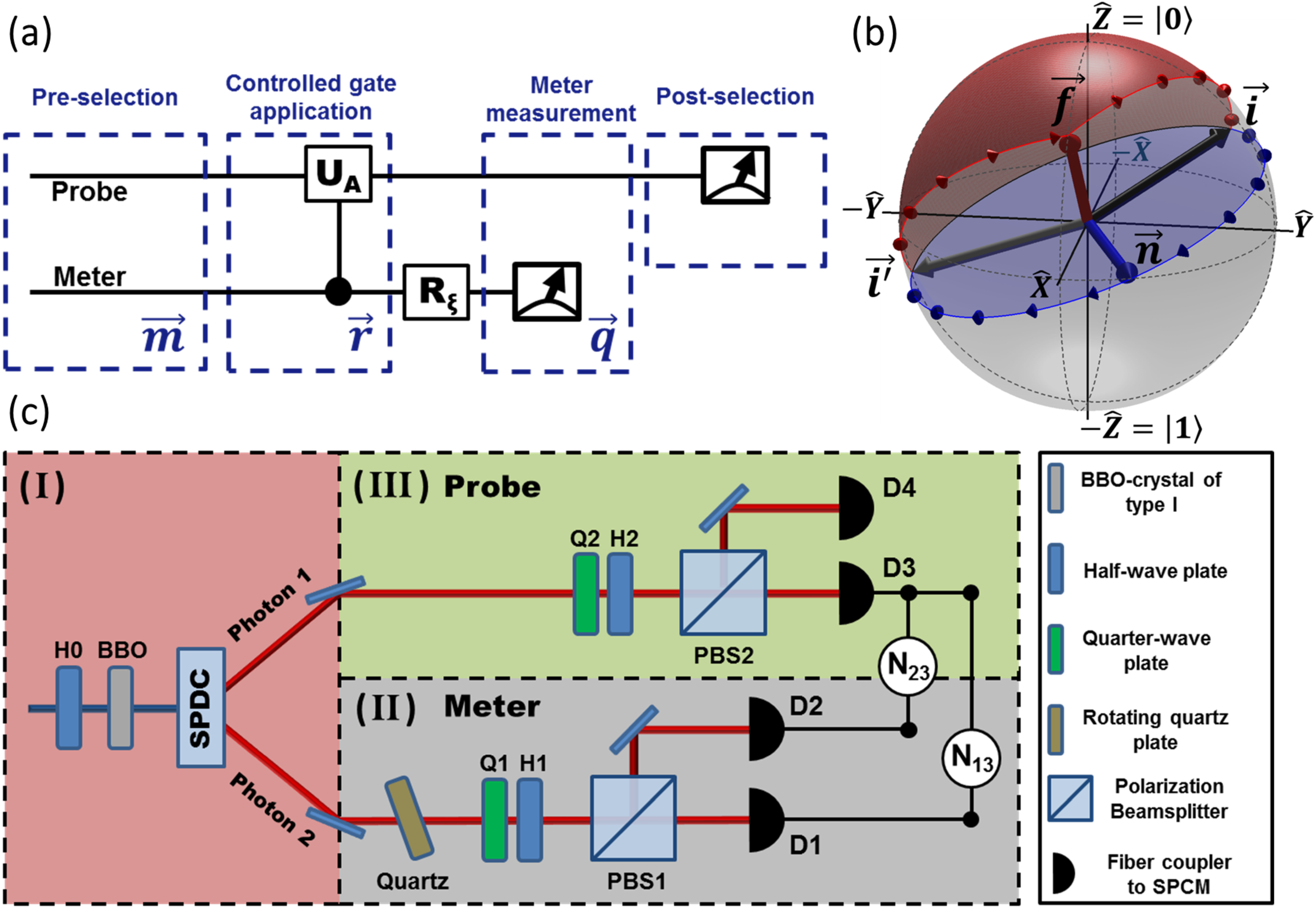} 
            \caption{(color online) Quantum controlled evolution: (a) protocol, (b) representation in the Bloch sphere of the relevant  probe states and (c) experimental set-up. (b)  Probe Bloch sphere with initial $\vec{i}$ and final $\vec{f}$ states, $\hat{\sigma}_n$ observable rotation axis $\vec{n}$, and the mirror image $\vec{i}^\prime$ of $\vec{i}$ with respect to the $\vec{n}$ axis. The solid angle  $\Omega_{i n i^\prime\! f}$  associated to the geometric phase in eq. (\protect\ref{eq:GmPhase}) is obtained by following consecutively the three great circle arcs $i \rightarrow n \rightarrow i^\prime$ (blue), $i^\prime \rightarrow f$ (red), and $f \rightarrow i$ (red). (c) The set-up comprises three areas: the state preparation with the two qubit generation (\mbox{I}), the meter measurement by detectors $D_1$ and $D_2$ (\mbox{II}) and the final probe post-selection by $D_3$ (\mbox{III}). The coincidence counts $N_{13}$ and $N_{23}$ are acquired by four single photon counting modules (SPCM) placed in the meter and probe paths.} 
            \label{fig:bloch representation}
	\end{center}
\end{figure} 

Now we consider the connection between modular and weak values to gain insight into the physics of weak values. The previously arbitrary probe system becomes a qubit and the probe transformation $\hat{U}_{A}=e^{-i\frac{g}{2}\,\hat{\sigma}_{n}}$ is a rotation operator involving the Pauli observable $\hat{\sigma}_{n}=\vec{n}.\hat{\vec{\sigma}}$ ($\vec{n}$ a unit vector). We set a strong AAV coupling strength $g=\pi$. Then, $\hat{U}_{A}=-i\,\hat{\sigma}_{n}$ and the quantum gate  acting on the two qubits becomes:
\begin{equation}
\hat{U}_{GATE}=\hat{\Pi}_{r}\otimes\hat{I}+\hat{\Pi}_{-r}\otimes\hat{\sigma}_{n}\,,
\end{equation}
where the phase factor $\delta$ in (\ref{eq:controlled measurement interaction}) was set to $\frac{\pi}{2}$. This shows the equivalence of modular and weak values of $\hat{\sigma}_{n}$ (see also \cite{Kedem (2010)}). We can thus apply our scheme to determine an arbitrary weak value of the Pauli operator in its polar representation. Interestingly, the argument of the weak value depends only on the probe evolution from its initial to final state as defined by the operator $\hat{U}_A$ (a related result is mentioned in \cite{Pati (2014)}). This phase has a topological component, similar to the Pancharatnam geometric phase. Its value is proportional to the solid angle $\Omega_{i n i^\prime\! f}$ delimited by four vectors on the Bloch sphere (see Fig. \ref{fig:bloch representation}.b and derivation \cite{Supplemental Material Geometric Phase} for details):
\begin{equation}
\arg \frac{\langle\psi_f | \hat{ \sigma}_n | \psi_i \rangle}{\langle \psi_f | \psi_i \rangle}=\arctan\frac{(\vec{n}\times\vec{i}\,).\vec{f}}{\vec{n}.\vec{i}+\vec{n}.\vec{f}}=-\frac{1}{2} \Omega_{i n i^\prime\! f}\,.\label{eq:GmPhase}
\end{equation}
This geometric phase is completely defined by intrinsic properties of the probe system. It is observed directly in the interferometric experiment but does not depend on the meter properties, contrary to the pointer shift in usual weak measurement (which depends on $g$). It emphasizes the quantum origin of the argument of the complex weak value and its relationship to physical properties of the probe system.

Experimentally, we implement a conceptual CNOT gate $\hat{U}_{GATE}=\hat{\Pi}_{|0\rangle}\otimes\hat{I}+\hat{\Pi}_{|1\rangle}\otimes\hat{\sigma}_{x}$. The initial meter state $\hat{\rho}_{m}=\frac{1}{2} \hat{I}+\frac{1}{2}P_{m}\,\overrightarrow{m}.\hat{\overrightarrow{\sigma}} $, with $\overrightarrow{m} =\left(\sin\theta,\,0,\, \cos\theta\right)$, controls the application of the unitary observable $\hat{\sigma}_{x}$ on the target probe state pre-selected in the $|\psi_{i}\rangle=|0\rangle$ state. The meter projective measurement is then performed in the $\hat{\sigma}_{x}$ basis ($\vec{q}$). It erases the information about the application of $\hat{\sigma}_{x}$ on the target since it was controlled by the meter basis vectors $|0\rangle$ and $|1\rangle$ of $\hat{\sigma}_{z}$ ($\vec{r}$). It is followed by the probe measurement of the final post-selected state $|\psi_{f}\rangle$. Finally, we obtain the weak value $\sigma_{x,w}$ as a function of our chosen initial meter state ($\vec{m}$), which defines the measurement strength ($\theta$), and post-selected probe state  $|\psi_{f}\rangle$.

In practice, the two-qubit state after the CNOT gate is simulated by entangled photon pairs produced by type I spontaneous parametric down-conversion in two orthogonal nonlinear BBO-crystals  \cite{Kwiat (1999)} (see Fig. \ref{fig:bloch representation}.c and \cite{Supplemental Material Experimental Material} for set-up details). One photon is assimilated to the meter and the other to the probe. A half-wave plate and a third BBO placed before the source control the photon pair state. Conceptually, they select the initial meter polarization of the CNOT gate while the initial probed state always has horizontal polarization. 
\begin{figure}[t]
\centering{}\includegraphics[scale=0.24]{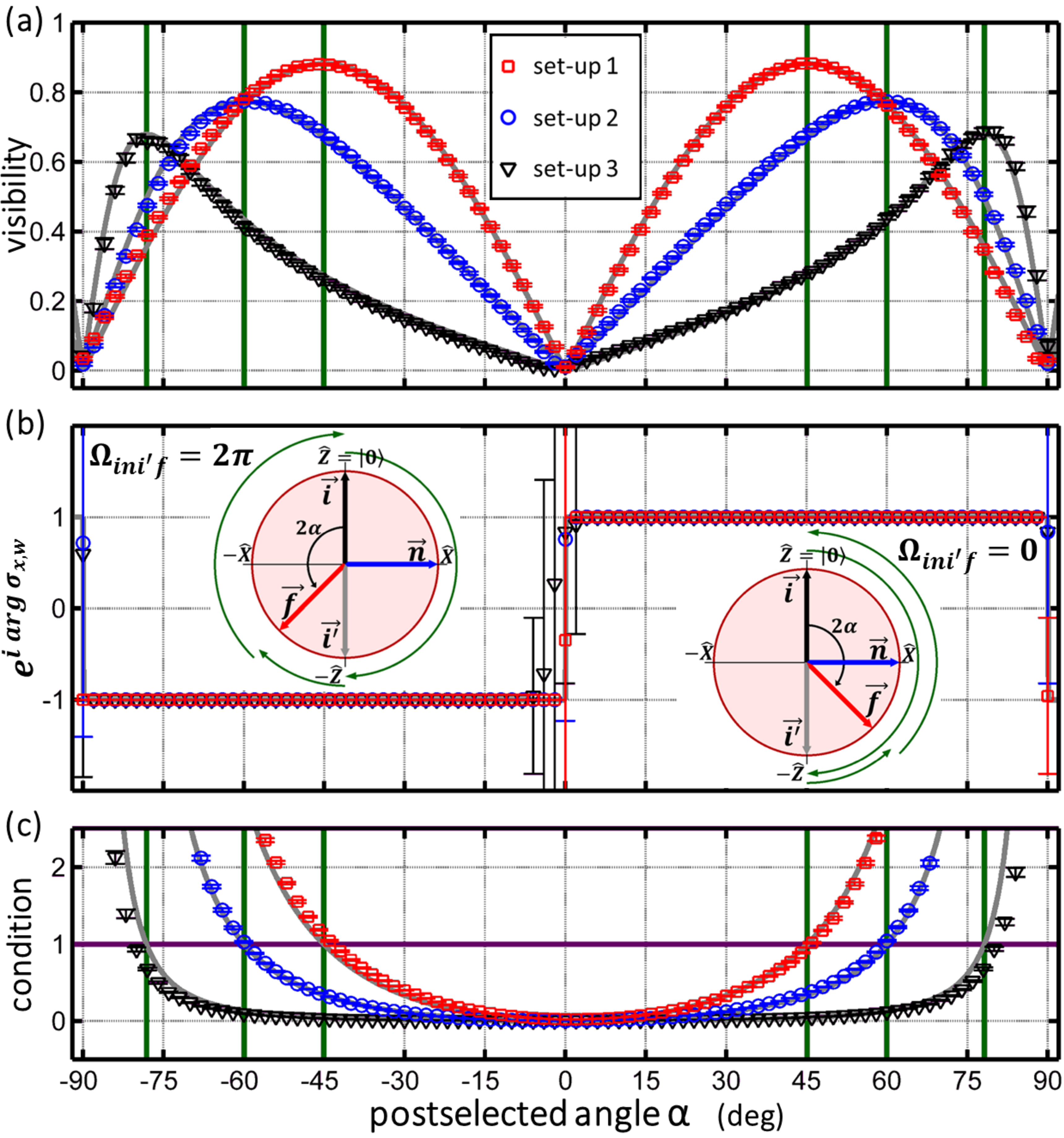}\caption{(color online) (a) Visibility and (b) argument  as a function of the post-selected polarization $|\psi_{f}\rangle= \cos \left(\alpha\right)\,|H\rangle+\sin\left(\alpha\right)|V\rangle$ with pre-selected $|H\rangle$ polarization for three initial meter states: $\theta_{1} = 0.499\,\pi$ and $P_{1m} =0.882\pm0.002 $ (red squares), $\theta_{2} = 0.297\,\pi$ and $P_{2m} =0.836\pm0.002$ (blue circles),  $\theta_{3} =  0.092\,\pi$ and $P_{3m} =0.956\pm0.001$ (black triangles). (c) Positive/negative solution criterion.  Values larger/smaller than unity (violet, solid horizontal line) admit the positive/negative solution $\left|\sigma_{x, w}\right|_{\pm}$, respectively. Final meter states are $|D\rangle$  and $|A\rangle$ for measurements (a-b) and $|H\rangle$ and $|V\rangle$ for (c). Grey, solid lines represent theoretical curves.
\label{fig:VisibilityD}}
\end{figure}
We post-select the probe polarization $|\psi_{f}\rangle=cos\left(\alpha\right)\,|H\rangle+sin\left(\alpha\right)|V\rangle$ at detector $D_3$. Detectors $D_1$ and $D_2$ measure the meter polarization (diagonal $|D\rangle$ and anti-diagonal $|A\rangle$ states, respectively). We adjust the phase $\xi$ by tilting a birefringent $Z$-cut quartz plate in the meter path to obtain the interference visibility $V$ from the coincidence counts. When coincidence counts $N^{max}_{13}$ are maximal for detectors $D_1$ and $D_3$ (constructive interference), coincidence counts $N^{min}_{23}$ for detectors $D_2$ and $D_3$ are minimal (destructive interference). Then, the phase $\xi$  equals the argument $\varphi$ of the weak value $\sigma_{x,w}=\langle \psi_f(\alpha)|\hat{ \sigma}_x|H\rangle/\langle \psi_f(\alpha)|H\rangle$, while the visibility is given by equation (\ref{eq:visibility}): $V=\frac{N^{max}_{13}-N^{min}_{23}}{N^{max}_{13}+N^{min}_{23}}$. This scheme improves the signal-to-noise ratio of a weak measurement \cite{Supplemental Material SNR}.

Fig. \ref{fig:VisibilityD} presents the visibility and the phase for three different initial meter states inconciliable with weak measurement approximations. The corresponding strengths $\theta$ were determined from the density operator of the biphoton using quantum state tomography \cite{Altepeter (2004), Supplemental Material Preliminary Meter Analysis}. The purities were rather estimated by fitting the theoretical visibility to the data ($\chi^{2}$ minimization method) because $V$ is highly sensitive to $P_{m}$. The latter step may be skipped if the meter is supposed in a pure initial state \cite{Supplemental Material Preliminary Meter Analysis}, as usually done in the literature. We chose pre- and post-selected states for which the argument of $\sigma_{x,w}(\alpha)$ (Fig. \ref{fig:VisibilityD}.b) takes only the two values $0$ or $\pi$ for simplicity. It determines the sign of the weak value. The solid angle $\Omega_{i n i^\prime\! f}$ related to the geometric phase (\ref{eq:GmPhase}) is defined in the $OXZ$ plane of the Bloch sphere (see Fig. \ref{fig:VisibilityD}.b): the great circle arcs $i \rightarrow n \rightarrow i^\prime \rightarrow f \rightarrow i$ make a full circle or compensate each other depending on the post-selected state $f(\alpha)$. The visibility (Fig. \ref{fig:VisibilityD}.a) provides the modulus $\left|\sigma_{x,w}\right|$, using the two solutions obtained in relation (\ref{eq:modular value relation}). The switch between them occurs at the maximum of the visibility. The criterion (\ref{eq:WeakMeasurCond}) determining this switch is measured from the coincidence count ratio $N^{c}_{23}/N^{c}_{13}$ with horizontal $|H\rangle$ (detector $D_1$) and vertical $|V\rangle$ (detector $D_2$) meter polarizations. In this case $\vec{r}=\vec{q}$, and the meter measurement reveals completely the probe state after the quantum gate interaction (no information erasure). The theoretical switch angle and the measured criterion agree strongly except for strengths approaching the range of weak measurements ($\theta_3$), where a difference of $2-4^\circ$ is observed due to increasing experimental noise (Fig. \ref{fig:VisibilityD}.c).

\begin{figure}[t]
\centering{}\includegraphics[scale=0.225]{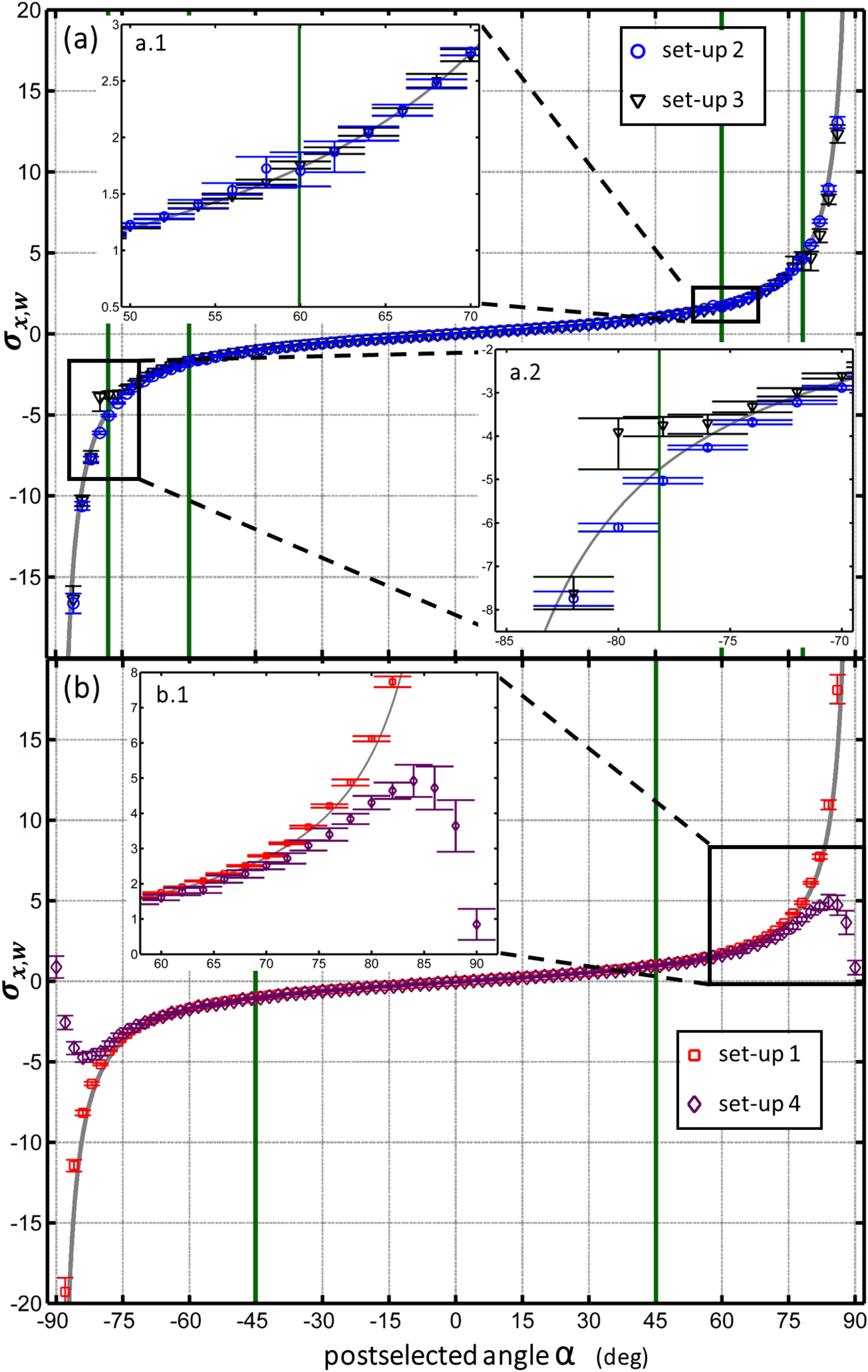}\caption{(color online) Weak values determined from phase and visibility measurement for the three strengths $\theta_1$ (red squares), $\theta_2$ (blue circles), and $\theta_3$ (black triangles), and from the standard weak measurement technique (violet diamonds) using relation (\ref{eq:realpartmodular variable}) with the weak strength $\theta_{4}= 0.025\,\pi$ and purity $P_{4m} =0.982\pm0.001$. All set-ups use a $|D\rangle, |A\rangle$ basis for the final meter measurement (the additional measurement in the $|L\rangle, |R\rangle$ meter basis required by standard weak measurements was not performed since $\Im \mathfrak{m}\, \sigma_{x,w}(\alpha)=0$ here).  \label{fig:WV}}
\end{figure}
The full weak values determined using a strong ($\theta_2$) or a weaker ($\theta_3$) strength are compared in Fig. \ref{fig:WV}.a. Both set-ups provide excellent agreement with the theoretical curve, except at the solution switch, where the accuracy of the set-up using weaker measurement strengths decreases (see insets a.1 and a.2). In Fig. \ref{fig:WV}.b, we compare our method to the standard weak measurement technique. For a small modulus of the weak value $\sigma_{x,w}(\alpha)$, both techniques provide results close to theoretical predictions. However, for large moduli, the weak measurement approximation breaks down completely (zoom b.1) for a wide range of post-selected states approaching orthogonality to the pre-selected state. Weak measurement results are useless there and only our method works.

In conclusion, the presented quantum eraser procedure exploits a qubit meter to measure directly the modulus and the argument of complex modular and weak values for arbitrary measurement strengths. The connection between modular and weak values allowed us to investigate directly weak values of qubit systems in their polar representation by performing a one-step visibility and phase measurement. In this case, the argument of the weak value is associated to a quantum geometric phase, that has a non-controversial physical meaning. This direct relevance of the polar form of the weak value to the intrinsic physical properties of the system evolution shows that the interpretation of past and present experiments involving weak values ought not be limited to the consideration of their real and imaginary parts. Our method to determine weak values requires fewer measurements and does not suffer the limitations of the standard weak measurement technique for large weak values, while it is applicable for both weak and strong measurement conditions. Experimentally, this opens the way to exploiting with greater accuracy the measure of weak values, particularly for nearly orthogonal pre- and post-selected states.
 
\begin{acknowledgments}
Y. C. is a research associate of the Belgian Fund for Scientific Research F.R.S.-FNRS. B. K. acknowledges financial support from the Action de Recherche Concert\'{e}e (BIOSTRUCT project) of the University of Namur (UNamur) and the support from the Nanoscale Quantum Optics COST-MP1403 action. The authors would like to thank Profs. B. Hespel and P. A. Thiry for fruitful discussions and support, without which this work would not have been possible. We also thank S. Mouchet for the careful reading of the paper and the insightful remarks as well as J.-P. van Roy for the upgrade and the automation of the whole photon detection system.
\end{acknowledgments}

\section{Supplemental Material}

\subsection{Meter configurations}

In our quantum protocol, the average $\overline{\sigma}_{q}^{m}$ of the meter observable for a given pre- and post-selected sub-ensemble of the probe system is:
\begin{equation}
\overline{\sigma}_{q}^{m}=2 P_{m}\frac{\left(\overrightarrow{m}.\overrightarrow{q}\right)\: \operatorname{\Re \mathfrak{e}} A_{m} +\left[ \left(\overrightarrow{r}\times\overrightarrow{m}\right).\overrightarrow{q}\right] \:\operatorname{\Im \mathfrak{m}} A_{m} }{\left(1+P_{m}\,\overrightarrow{r}.\overrightarrow{m}\right)\:+\left(1-P_{m}\, \overrightarrow{r} . \overrightarrow{m}\right)\: \left|A_{m}\right|^{2}}\:.\label{eq:QSMeanValue}
\end{equation}
For a given vector $\vec{r}$ controlling the quantum gate action, the quantum eraser condition $\overrightarrow{r}.\overrightarrow{q}=0$ constrains $\vec{q}$ to the red plane in figure \ref{fig:bloch representation}.a. We choose now particular final vectors $\vec{q}$ of the meter system in relationship to the initial vector $\vec{m}$ (characterizing the meter initial state), in order to determine the real and imaginary parts of the modular value from the average meter observable $\overline{\sigma}_{q}^{m}$. We pick the real part of $A_{m}$ when the three vectors $\vec{m},\, \vec{r},\, \vec{q}$ are coplanar ($\vec{q}_{Re}$ in blue plane in figure \ref{fig:bloch representation}.b), so that $\left(\overrightarrow{r}\times\overrightarrow{m}\right).\overrightarrow{q}=0$ in equation (\ref{eq:QSMeanValue}). We isolate the imaginary part with orthogonal initial and final states of the meter ($\vec{q}_{Im}$ orthogonal to blue plane in figure \ref{fig:bloch representation}.c), so that $\overrightarrow{m}.\overrightarrow{q}=0$ in equation (\ref{eq:QSMeanValue}). 

\begin{figure}[b]
	\begin{center}
			\includegraphics[width=0.48\textwidth]{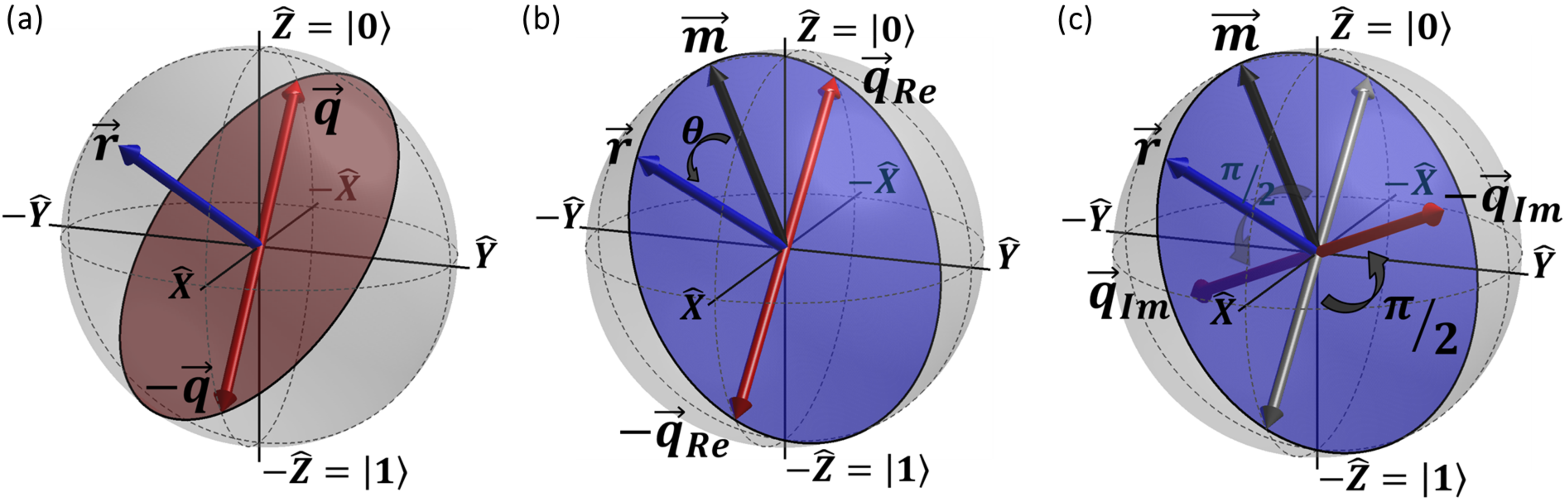} 
            \caption{Representation in the Bloch sphere of the relevant (a-c) meter states. (a) The red plane is perpendicular to the control state $\protect\overrightarrow{r}$. It contains all final meter states $\protect\overrightarrow{q}$ and $-\protect\overrightarrow{q}$ implementing the quantum eraser condition. (b-c) The blue plane contains the initial meter state $\protect\overrightarrow{m}$ and the control state $\protect\overrightarrow{r}$. The final meter states (b) $\protect\overrightarrow{q}\!_{Re}$ in the blue plane and (c) $\protect\overrightarrow{q}\!_{Im}$ perpendicular to it measure the real and imaginary parts of the modular value, respectively.} 
            \label{fig:bloch representation}
	\end{center}
\end{figure}

\subsection{Topological component of the weak value argument}

As described in the letter, the argument of the weak value of the spin operator $\hat{\sigma}_{n}$ verifies the two equalities:
\begin{equation}
\arg \frac{\langle f | \hat{ \sigma}_n | i \rangle}{\langle f | i \rangle}=\arctan\frac{(\vec{n}\times\vec{i}\,).\vec{f}}{\vec{n}.\vec{i}+\vec{n}.\vec{f}}=-\frac{1}{2} \Omega_{i n i^\prime\! f}\,.\label{eq:GmPhase}
\end{equation}

The first equality results immediately from the definition of the argument of the weak value, considering the following expression:   
\begin{equation}
\arg \frac{\langle f | \hat{ \sigma}_n | i \rangle}{\langle f | i \rangle}= \arctan\frac{\operatorname{\Im \mathfrak{m}}\,\langle f|\hat{ \sigma}_n|i\rangle \langle i | f \rangle}{\operatorname{\Re \mathfrak{e}}\,\langle f|\hat{ \sigma}_n|i\rangle \langle i | f \rangle}\,,
\end{equation}
in which  $\langle f|\hat{ \sigma}_n|i\rangle \langle i|f\rangle $ can be expressed as a function of directions on the Bloch sphere:
\begin{equation}
\langle f|\hat{ \sigma}_n|i\rangle \langle i|f\rangle=\frac{1}{2} (\,\vec{n}.\vec{i}+\vec{n}.\vec{f}+j\,(\vec{n}\times\vec{i}\,).\vec{f}\,)\,, 
\end{equation}
where $j$ is the imaginary unit, $\vec{i}$ and $\vec{f}$ describe the pure initial and final qubit states, respectively, and $\vec{n}$ gives the rotation axis associated to the observable $\hat{\vec{\sigma}}_n$. For the second equality in (\ref{eq:GmPhase}), our approach is inspired by the work of Martinez et al. \cite{Martinez (2012)}, where they studied the geometrical characteristics (amplitude and phase) of polarization modulation optical devices on the Poincar\'{e} sphere.

The trajectory of a pure qubit state on the Bloch sphere corresponding to the unitary transformation $\hat{\sigma}_{n}$ is a non-geodesic opened arc. Consequently, the resulting state $|I^\prime\!\rangle=\hat{\sigma}_{n}|i\rangle$ is no longer in phase with the initial state $|i\rangle$. To yield nonetheless an expression in terms of a solid angle for the accumulated phase, let us express the resulting state as $|I^\prime\!\rangle=e^{j\varphi_{i\rightarrow i^\prime\!}}|i^\prime\!\rangle$, where the phase $\varphi_{i\rightarrow i^\prime\!}$ is due to the non-geodesic  movement of $|i\rangle$ to the output state $|i^\prime\!\rangle$. The vector $\vec{i}^{\prime}$ is entirely defined by $\vec{i}^{\prime}=2\left(\vec{n}.\vec{i}\right)\,\vec{n}-\vec{i}$ (essentially, $\vec{i}^{\prime}$ is the mirror image of $\vec{i} $ with respect to the $\vec{n}$ axis). The additional phase $\varphi_{i\rightarrow i^\prime\!}$ is determined by projecting $|I^\prime\!\rangle$ onto the orthogonal eigenvectors $|n\rangle$ and $|-n\rangle$ of the operator $\hat{\sigma}_{n}$. By considering:
\begin{equation}
\hat{\sigma}_{n}=|n\rangle\langle n|-|-n\rangle\langle-n|\,,
\end{equation}
we conclude that the projection of $|I^\prime\!\rangle$ onto the eigenvector $|n\rangle$ yields the following two relations: 
\begin{equation}\label{connectionRel}
\langle n|I^\prime\!\rangle	=	e^{j\varphi_{i\rightarrow i^\prime\!}}\langle n|i^\prime\!\rangle
=\langle n|i\rangle\,.
\end{equation}
The moduli $\left|\langle n|i\rangle\right|=\left|\langle n|i'\rangle\right|$ are equal since $\vec{i}$ and $\vec{i^\prime}$ are mirror images with respect to $\vec{n}$. Consequently, only the accumulated total phase of the open loops, known as Pancharatnam connection, remain in (\ref{connectionRel}):
\begin{equation}\label{phiexpr}
\varphi_{i\rightarrow i^{^{\prime}}} = \arg\langle n|i\rangle-\arg\langle n|i'\rangle\:.
\end{equation}

\begin{figure}[t]
\centering\includegraphics[scale=0.12]{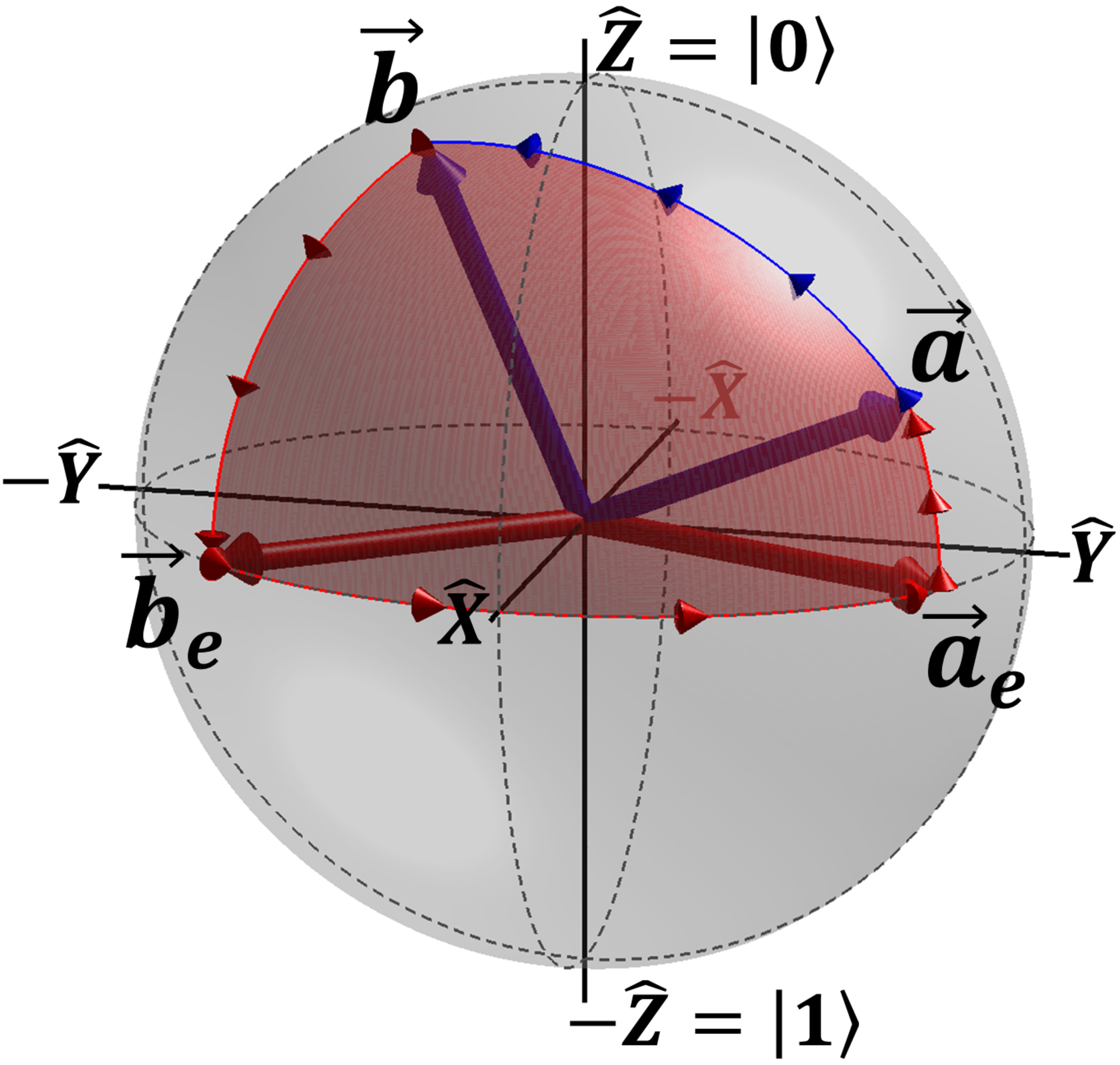}
\caption{The solid angle $\Omega_{abb_{e}a_{e}}$ of the four vertices $\vec{a}$, $\vec{b}$, $\vec{b}_{e}$ and $\vec{a}_{e}$ represented in the Bloch sphere (the red surface). The states $|a_{e}\rangle$ and $|b_{e}\rangle$ lying on the equator are horizontal lifts of the associated states $|a\rangle$ and $|b\rangle$, respectively. The solid angle of the counterclockwise sequence of  states $|a\rangle\rightarrow|b\rangle\rightarrow|b_{e}\rangle\rightarrow|a_{e}\rangle\rightarrow|a\rangle$ is in direct relationship to the Pancharatnam connection $\arg\ \langle b|a\rangle$.\label{fig:sphericalquadrangles}}
\end{figure}

In practice, the Pancharatnam connection $\arg\langle b|a\rangle$ relating arbitrary states $|a\rangle$ and $|b\rangle$ is determined by calculating the spherical quadrangle $\Omega_{abb_{e}a_{e}}$ in the Bloch sphere (figure \ref{fig:sphericalquadrangles}), where the supplemental vertices $|a_{e}\rangle$ and $|b_{e}\rangle$ are well-defined vectors. To understand, how they are determined, we must express their position in the spherical coordinate system. By convention, $2\eta$ corresponds to the azimuth angle and $2\chi$ to the polar angle. In this representation, a pure state on the Bloch sphere is defined by: 
\begin{equation}
\vec{a}\left(\chi,\eta\right)=\left(\begin{array}{c}\cos\left(2\eta\right)\, \cos(2\chi)\\ \sin\left(2\eta\right)\, \cos(2\chi)\\ \sin(2\chi)
\end{array}\right)\:,
\end{equation}
and the Pancharatnam connection is given by:
\begin{equation}
\arg\langle b|a\rangle=\arctan\left(\tan\left(\eta_{a}-\eta_{b}\right)\frac{\sin\left(\chi_{a}+\chi_{b}\right)}{\cos\left(\chi_{a}-\chi_{b}\right)}\right).
\end{equation}
The connection is in phase, i.e. $\arg\langle b|a\rangle=0$, for transports with the same azimuth angles $\eta$ and for transformations happening around the equator of the Bloch sphere, i.e. for the polar angle $\chi=0$. These two kinds of transports are known as horizontal lifts along the geodesic connecting the states $|a\rangle$ and $|b\rangle$ on the Bloch sphere. The states $|a_{e}\rangle=|\eta_{a},0\rangle$ and $|b_{e}\rangle=|\eta_{b},0\rangle$ are fixed with the same azimuth angle as $|a\rangle$ and $|b\rangle$, respectively, and with a polar angle $\chi=0$.

The closed loop $|a\rangle\rightarrow|b\rangle\rightarrow|b_{e}\rangle\rightarrow|a_{e}\rangle\rightarrow|a\rangle$ along the geodesic arcs determines the spherical quadrangle $\Omega_{abb_{e}a_{e}}$, which is equivalent to $\arg\langle b|a\rangle$:
\begin{eqnarray}
\arg\langle b|a\rangle & = & \arg\langle a|a_{e}\rangle+\arg\langle a_{e}|b_{e}\rangle+\arg\langle b_{e}|b\rangle+\arg\langle b|a\rangle\nonumber \\
 & = & \arg\left(\langle a|a_{e}\rangle\langle a_{e}|b_{e}\rangle\langle b_{e}|b\rangle\langle b|a\rangle\right)\nonumber \\
 & = & -\frac{\Omega_{abb_{e}a_{e}}}{2}\:.\label{eq:argsolidabba}
\end{eqnarray} 
Note that the sign present in front of the solid angle for a given sequence of states is positive when the sequence is followed counterclockwise and is negative when the sequence is followed clockwise. The sign of the solid angle changes when the sequence of projections is inversed, $\Omega_{a\rightarrow b}=-\,\Omega_{b\rightarrow a}$. It is possible to express a solid angle linking three vertices as a sum of three spherical quadrangles \cite{Martinez (2012)}:
\begin{equation} \label{decomp}
\Omega_{abc}=\Omega_{abb_{e}a_{e}}+\Omega_{bcc_{e}b_{e}}+\Omega_{caa_{e}c_{e}}\: ,
\end{equation}
where each spherical quadrangle contains two vertices of the initial solid angle.

We use the decomposition property of eq. (\ref{decomp}) to rewrite the expression giving $\varphi_{i\rightarrow i^{\prime}}$ (\ref{phiexpr}) according to: 
\begin{equation}\label{triquadforphi}
\varphi_{i\rightarrow i^{\prime}}=-\frac{\Omega_{ini^{\prime}}+\Omega_{ii^{\prime}i'_{e}i_{e}}}{2}\:,
\end{equation}
where we made use of eq. (\ref{eq:argsolidabba}) to express the connexions appearing in (\ref{phiexpr}). Following the indices may be tedious but, essentially, eq. (\ref{phiexpr}) and (\ref{eq:argsolidabba})  show together that the expression of the phase $\varphi_{i\rightarrow i^{\prime}}$ includes a sum of two spherical quadrangles; then we used eq. (\ref{decomp}) to express the sum of these two spherical quadrangles as a function of the third spherical quadrangle and of the  spherical triangle appearing in eq. (\ref{decomp}).

Furthermore, expression (\ref{triquadforphi}) points out that the non-geodesic phase $\varphi_{i\rightarrow i^{\prime}}$
is the sum of the geometric phase of the closed loop $|i\rangle\rightarrow|n\rangle\rightarrow|i'\rangle\rightarrow|i\rangle$ (first term)
and the phase of the Pancharatnam connection $|i\rangle\rightarrow|i^{\prime}\rangle$ (second term).

Using the last results, the argument of the weak value of $\hat{\sigma}_{n}$
is: 
\begin{eqnarray}
\arg\frac{\langle f|\hat{\sigma}_{n}|i\rangle}{\langle f|i\rangle} & \stackrel{(a)}{=} & \arg\left(e^{j\varphi_{i\rightarrow i^{\prime}}}\frac{\langle f|i^{\prime}\rangle}{\langle f|i\rangle}\right)\nonumber \\
 & \stackrel{(b)}{=} & \arg\left(\frac{\left| \langle f|i^\prime\rangle  \right|}{\left|  \langle f|i\rangle \right|}e^{j\varphi_{i\rightarrow i^{\prime}}}e^{-j\frac{\Omega_{i'ff_{e}i'_{e}}}{2}}e^{-j\frac{\Omega_{fii_{e}f_{e}}}{2}}\right)\nonumber \\
 & \stackrel{(c)}{=} & \varphi_{i\rightarrow i'}-\frac{\Omega_{i'ff_{e}i'_{e}}+\Omega_{fii_{e}f_{e}}}{2}\nonumber \\
 &\stackrel{(d)}{=} & \varphi_{i\rightarrow i'}-\frac{\Omega_{ii'f}-\Omega_{ii'i'_{e}i_{e}}}{2}\nonumber \\
 & \stackrel{(e)}{=} & -\frac{\Omega_{ini'}}{2}-\frac{\Omega_{ii'i'_{e}i_{e}}}{2}-\frac{\Omega_{ii'f}}{2}+\frac{\Omega_{ii'i'_{e}i_{e}}}{2}\nonumber \\
 & \stackrel{(f)}{=} & -\frac{\Omega_{ini'}}{2}-\frac{\Omega_{ii'f}}{2}\nonumber \\
 & \stackrel{(g)}{=} & -\frac{\Omega_{ini'f}}{2}\:.
\end{eqnarray}
Equality (a) results from the definition of the states $|I^\prime\rangle$ and $|i^\prime\rangle$. (b) expresses the Pancharatnam connexions in terms of solid angles using eq. (\ref{eq:argsolidabba}). (c) takes the argument of the previous expression. (d) exploits the decomposition property of eq. (\ref{decomp}). (e) follows from the expression of $\varphi_{i\rightarrow i^{\prime}}$ in (\ref{triquadforphi}). (f) is due to canceling terms. (g) combines the two spherical triangles in one spherical quadrangle (as the paths $i\rightarrow i^\prime$ and $i^\prime\rightarrow i$ present in the triangles cancel each other).

\subsection{Experimental material} 

Experimentally, we implement the polarization-entangled biphoton state after the conceptual CNOT gate using two orthogonal nonlinear BBO crystals in the ``sandwich configuration" \cite{Kwiat (1999)}. Each crystal is $1$ mm thick and cut for type-I phase matching with $\theta=29.2^{o}$ and $\varphi=90^{o}$. Via the nonlinear interaction of spontaneous parametric down-conversion (SPDC), the pump laser (blue diode DL-7146-101S from SANYO Electric Co.) centered at $407$ nm  generates two polarization-entangled photons at $814$ nm. The laser diode is controlled by temperature (Thorlabs TED 200C) and current (Thorlabs LDC202C) controllers. It produces a continuous laser output power of $60$ mW. We use four single photon counting modules (SPCM-AQ4C from Perkin-ElmerFor) for the joint polarization measurement of the meter or probe photons. The polarization basis are selected by half- and quarter-wave plates followed by a polarizing beam-splitter (RCHP-15.0-CA-670-1064 from CVI Melles Griot). Before detection, the photons are coupled into multimode fibers and filtered using low-pass filters (FGL780 from Thorlabs). About $4000$ total coincidence counts per second are acquired by using a homemade FPGA (SE3BOARD from Xilinx) coincidence counter.

\subsection{Signal-to-noise ratio} 

By definition, the signal-to-noise ratio (SNR) is the ratio of the magnitude of the expected meter shift to the standard deviation \cite{Kofman (2012)}.

For given pre- and post-selected probe states, we consider a final meter outcome which follows a binomial distribution: the meter qubit is measured either by detector $D_{1}$ (and contributes to $N^{max}_{13}$) or by detector $D_{2}$ (and contributes to $N^{min}_{23}$). The expectation value of $N^{min}_{23}$  is therefore $E\left[N^{min}_{23}\right]=p\left(2|3\right)N$ and its variance $\Delta N^{min}_{23}=p\left(2|3\right)\left(1-p\left(2|3\right)\right)N$. $N=N^{max}_{13}+N^{min}_{23}$ is the total number of pre- and post-selected detector events and $p\left(2|3\right)$ is the conditional probability to trigger the meter detector $D_{2}$ for a given probe detection by $D_{3}$. Consequently, the expectation value of the measured visibility is:
\begin{eqnarray}
E_{V} & {=} & E\left[\frac{N^{max}_{13}-N^{min}_{23}}{N^{max}_{13}+N^{min}_{23}}\right]\nonumber \\
 & {=} & 1 - \frac{2}{N} E\left[N^{min}_{23}\right]\nonumber\\
 & {=} & 1 - 2 p\left(2|3\right)\nonumber  \\
 &  {=} & V\:,
\end{eqnarray} 
where the last equality follows directly from the definition of the conditional probability: $p\left(2|3\right)=\frac{P_{min}}{P_{max}+P_{min}}$, with $P_{max}$ and $P_{min}$ the maximum and the minimum of the joint probability of the measurement protocol. The corresponding variance is:
\begin{eqnarray}
\Delta_{V} & {=} & \Delta\left(\frac{N^{max}_{13}-N^{min}_{23}}{N^{max}_{13}+N^{min}_{23}}\right)\nonumber \\
 & {=} & \frac{4}{N^2} \Delta\left(N^{min}_{23}\right)\nonumber\\
 & {=} & \frac{4 p\left(2|3\right)\left(1-p\left(2|3\right)\right)}{N}\nonumber  \\
 &  {=} &\frac{1 - V^{2}}{N}\:,
\end{eqnarray} 
where we used the relationship $p\left(2|3\right)=\frac{1-V}{2}$. This leads to the standard deviation:
\begin{equation}\label{standarddeviation}
\sigma_{V}=\sqrt{\frac{1 - V^{2}}{N}}\:.
\end{equation}
Finally, the signal-to-noise ratio of the presented measurement scheme is:
\begin{equation}\label{SNR}
SNR=\frac{V}{\sqrt{1 - V^{2}}} \sqrt{N}\:.
\end{equation} 
The standard protocol determines the real (or the imaginary) part of the modular value by measuring the meter observable $\hat{\sigma}_{q_{Re}}$ (or $\hat{\sigma}_{q_{Im}}$). In this case, the visibility $V$ in the signal-to-noise relation (\ref{SNR}) is replaced by the absolute value of the meter average $\overline{\sigma}^{m}_{q_{Re}}$ (or $\overline{\sigma}^{m}_{q_{Im}}$). In the weak measurement limit,  the latter is related to the modular value by the approximation $\overline{\sigma}^{m}_{q_{Re}}\approx \theta \, \Re \mathfrak{e}\, A_{m}$ (or $\overline{\sigma}^{m}_{q_{Im}}\approx \theta \, \Im \mathfrak{m}\, A_{m}$). This expression is similar to the one obtained for our scheme in the weak measurement limit, relating the visibility to the modulus of the modular value: $V\approx \theta \, \left|A_{m}\right|$. Because the modulus of a complex number is always larger or equal than its real and imaginary parts ($\left|A_{m}\right|\geq  \Re \mathfrak{e}\, A_{m}$ and $\left|A_{m}\right|\geq  \Im \mathfrak{m}\, A_{m}$), our scheme improves the signal-to-noise ratio of the weak measurement compared to the standard protocol. (And, additionaly, it works when the weak measurement approximation fails.)

\subsection{Preliminary meter analysis} 

As described in the main Letter, the proposed measurement scheme involves the determinations of two parameters: the measurement strength $\theta$ and the purity $P_m$. These requirements are not specific to our scheme. In all weak value measurement protocols, the determination of the measurement strength $\theta$ is a necessary and inevitable process which requires a separate acquisition. Our scheme does not perform better or worse than other schemes in the literature in that respect. The additional determination of the purity $P_m$ is only required because we considered the most general case of an initial incoherent state of the qubit meter system. Most of the literature assumes the meter to be in a known pure state to avoid this supplementary step. To determine the initial meter state in our protocol, it is only necessary to perform the quantum tomography of a single qubit. In pratice, in our experimental implementation, we simulated the CNOT gate by using spontaneous parametric down-conversion (instead of using a true CNOT gate with two separate entries that could be characterized independently). For this reason, in our experiment, we determined the purity by performing quantum tomography of the two-qubit state since the input meter state could not be measured directly. This two-qubit tomography is not at all required by the proposed protocol but appears only as a side-effect of the practical implementation of our demonstrative experiment.


\begin{thebibliography}{10}
\bibitem{Comment Hint}A simple transformation of $\hat{U}=e^{-i\int\hat{H}_{int}(t)\,dt}$ with  the interaction Hamiltonian $\hat{H}_{int}=g(t)\,\hat{\Pi}_{-r}\otimes\hat{A}$ reveals the similitude between the von Neumann measurement scheme and our gate application $\hat{U}_{GATE}$.
\end{thebibliography}

\end{document}